\documentclass[reprint,amsmath,amssymb,aps,twocolumn,superscriptaddress]{revtex4-1}
\usepackage{graphicx}
\usepackage{dcolumn}
\usepackage{bm}
\usepackage{float}
\usepackage{amssymb}
\usepackage{color}
\usepackage{multirow}
\usepackage{stmaryrd}

\begin{document}

\title{Rare earth engineering in RMn$_6$Sn$_6$ topological kagome magnets}

\author{Wenlong Ma}
 \email{These authors contributed equally to this work.}
\affiliation{International Center for Quantum Materials, School of Physics, Peking University, Beijing 100871, China}
\author{Xitong Xu}
 \email{These authors contributed equally to this work.}
\affiliation{International Center for Quantum Materials, School of Physics, Peking University, Beijing 100871, China}
\affiliation{Anhui Key Laboratory of Condensed Matter Physics at Extreme Conditions, High Magnetic Field Laboratory, HFIPS, Anhui, Chinese Academy of Sciences, Hefei 230031, P. R. China}
\author{Jia-Xin Yin}
 \email{These authors contributed equally to this work.}
\affiliation{Laboratory for Topological Quantum Matter and Advanced Spectroscopy (B7), Department of Physics, Princeton University, Princeton, NJ, USA}
\author{Hui Yang}
\affiliation{International Center for Quantum Materials, School of Physics, Peking University, Beijing 100871, China}
\author{Huibin Zhou}
\affiliation{International Center for Quantum Materials, School of Physics, Peking University, Beijing 100871, China}
\author{Zi-Jia Cheng}
\affiliation{Laboratory for Topological Quantum Matter and Advanced Spectroscopy (B7), Department of Physics, Princeton University, Princeton, NJ, USA}
\author{Yuqing Huang}
\affiliation{International Center for Quantum Materials, School of Physics, Peking University, Beijing 100871, China}
\author{Zhe Qu}
\affiliation{Anhui Key Laboratory of Condensed Matter Physics at Extreme Conditions, High Magnetic Field Laboratory, HFIPS, Anhui, Chinese Academy of Sciences, Hefei 230031, P. R. China}
\affiliation{CAS Key Laboratory of Photovoltaic and Energy Conservation Materials, Hefei Institutes of Physical Sciences, Chinese Academy of Sciences, Hefei, Anhui 230031, China}
\author{Fa Wang}
\affiliation{International Center for Quantum Materials, School of Physics, Peking University, Beijing 100871, China}
\affiliation{Collaborative Innovation Center of Quantum Matter, Beijing 100871, China}
\author{M. Zahid Hasan}
\email{mzahid@princeton.edu}
\affiliation{Laboratory for Topological Quantum Matter and Advanced Spectroscopy (B7), Department of Physics, Princeton University, Princeton, NJ, USA}
\affiliation{Lawrence Berkeley National Laboratory, Berkeley, CA, USA}
\author{Shuang Jia}
 \email{gwljiashuang@pku.edu.cn}
\affiliation{International Center for Quantum Materials, School of Physics, Peking University, Beijing 100871, China}
\affiliation{Collaborative Innovation Center of Quantum Matter, Beijing 100871, China}
\affiliation{CAS Center for Excellence in Topological Quantum Computation, University of Chinese Academy of Sciences, Beijing 100190, China}
\affiliation{Beijing Academy of Quantum Information Sciences, West Building 3, No. 10 Xibeiwang East Road, Haidian District, Beijing 100193, China}
\date{\today}

\begin{abstract}
Exploration of the topological quantum materials with electron correlation is at the frontier of physics, as the strong interaction may give rise to new topological phases and transitions.
Here we report that a family of kagome magnets RMn$_6$Sn$_6$ manifest the quantum transport properties analogical to those in the quantum-limit Chern magnet TbMn$_6$Sn$_6$.
The topological transport in the family, including quantum oscillations with nontrivial Berry phase and large anomalous Hall effect arising from Berry curvature field, points to the existence of massive Dirac fermions.
Our observation demonstrates a close relationship between rare-earth magnetism and topological electron structure, indicating the rare-earth elements can effectively engineer the Chern quantum phase in kagome magnets.
\end{abstract}

\pacs{Valid PACS appear here}%

\maketitle

\begin{figure}[htbp]
	\begin{center}
		\includegraphics[clip, width=0.49\textwidth]{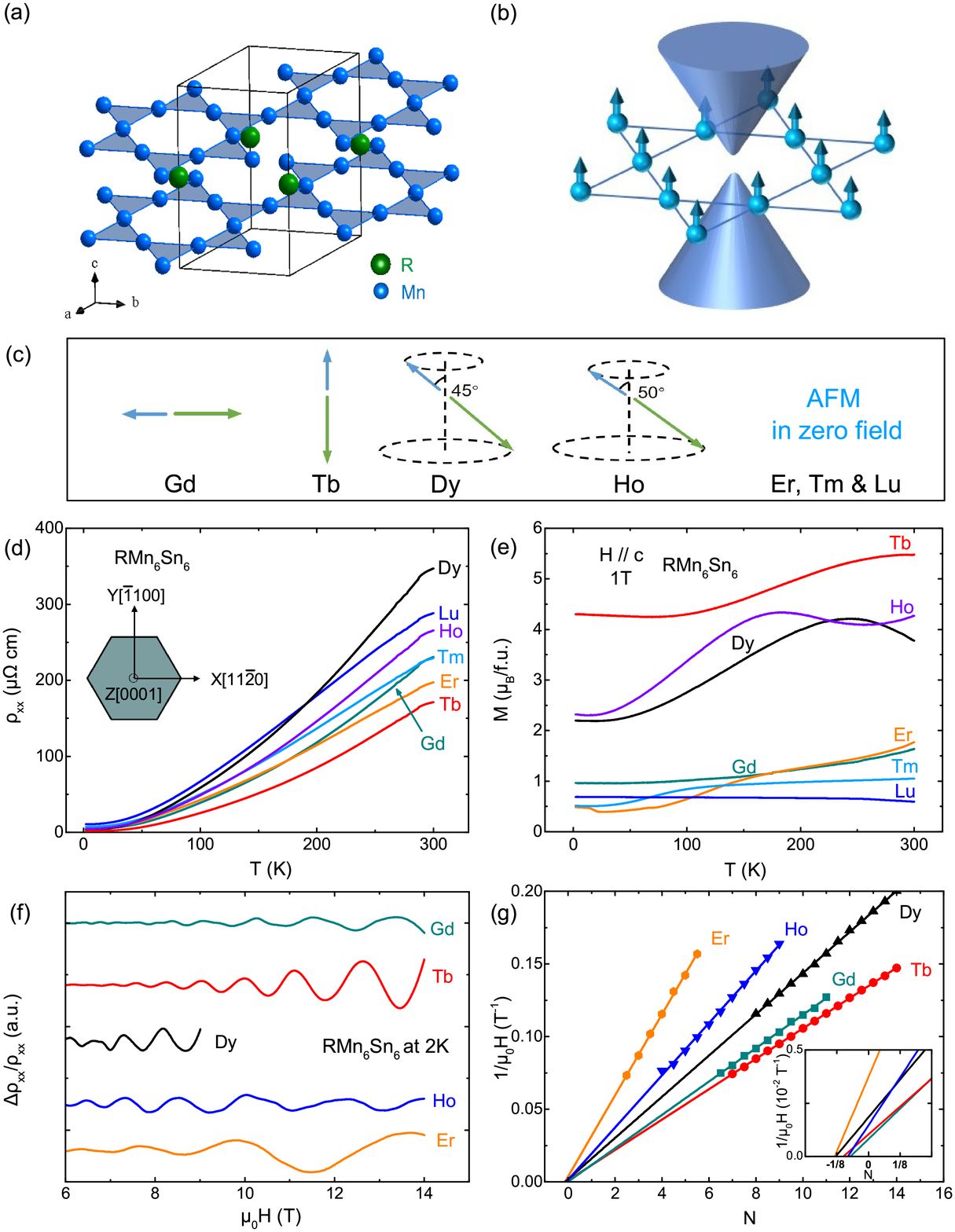}\\[1pt] 
		\caption{Crystal structure, physical property and topological electrons in RMn$_6$Sn$_6$.
(a) Mn kagome layers and R sublattices in RMn$_6$Sn$_6$.
(b) Illustration of MDF with a Chern gap in the magnetic kagome lattice.
(c) Magnetic structures in zero field. Blue and green arrows represent the direction of Mn and R moments, respectively.
(d) $\rho(T)$ curves for single crystals. Inset: Crystallographic orientations.
(e) M(T) for $\mu_0H=1 \mathrm{T}$.
(f) SdH QOs at 2 K.
(g) Landau fan diagram. Peak positions in (f) are assigned with half-integral Landau indexes~\cite{PhysRevLett.117.077201,murakawa2013detection}. Inset: Zoom-in near N = 0. }
		\label{fig:1}
	\end{center}
\end{figure}

The interplay between lattice geometry and electron correlation can create new states of quantum matter~\cite{keimer2017physics,sachdev2018topological,Canfield_2019,PhysRevLett.61.2015,chang2013experimental}, with recent examples including twisted bilayer graphene~\cite{Bistritzer12233,cao2018correlated}, Kitaev quantum spin liquid~\cite{KITAEV20062,kasahara2018majorana}, and kagome Chern magnet~\cite{PhysRevLett.115.186802,yin2020discovery}. A kagome lattice, consisting of two-dimensional corner-sharing triangles, naturally hosts both relativistic and dispersionless electrons~\cite{PhysRevB.62.R6065,zhang2011quantum}, which are the origins of its nontrivial band topology. With the inclusion of magnetism and spin-orbit coupling (SOC), kagome electrons can realize strongly interacting topological phases~\cite{PhysRevLett.115.186802,PhysRevB.62.R6065,PhysRevLett.106.236802}.
Recent study~\cite{yin2020discovery} has found a near-ideal quantum limit Chern magnet TbMn$_6$Sn$_6$, which hosts defect-free Mn kagome lattice.
However the topological nature of other members in the RMn$_6$Sn$_6$ (R = rare-earth element) family remains largely unexplored.
The RMn$_6$Sn$_6$ system features a pristine Mn$_3$ kagome lattice (Fig.~\ref{fig:1}(a)) in a layered structure~\cite{zhang2005unusual}.
They all exhibit magnetic ordering of Mn moments at room temperature while different R gives rise to various magnetic structures~\cite{VENTURINI199135,MALAMAN1999519,CLATTERBUCK199978}.
Of particular, TbMn$_6$Sn$_6$ manifests a ferrimagnetic (FIM) structure in which Tb sublattice is anti-parallel with out-of-plane ferromagnetically (FM) ordered Mn lattice.
This FIM structure has been proven to effectively sustain the spinless Haldane model generating Chern gapped massive Dirac fermions (MDFs), as illustrated in Fig.~\ref{fig:1}(b)~\cite{PhysRevLett.61.2015,PhysRevLett.115.186802,yin2020discovery}.

In this work we investigate the topological properties of the Mn kagome lattices in a series of RMn$_6$Sn$_6$ (R = Gd - Tm, Lu).
Their magnetic ground states are classified as FIM when R = Gd to Ho, and antiferromagnetic (AFM) when R is Er, Tm and Lu~\cite{zhang2005unusual} (Fig.~\ref{fig:1}(c)).
We found in general MDFs exist in the FM ordered kagome lattice of RMn$_6$Sn$_6$ but is absent in the AFM states.
The Chern gap ($\Delta$) and the Dirac cone energy ($E_D$) follow a decreasing trend from R = Gd to Er.
As we argue below, this R engineering on the topological electron structure is intimately related to the coupling between $4f$ local moment and Mn $3d$ moment.
In particular, $\Delta$ is proportional to the de Gennes factor ($dG$) of 4f moments while $E_D$ is proportional to $\sqrt{dG}$.
The interplay of the magnetic order and topological structure remains a fertile ground in the systems with strong correlation.
Our finding demonstrates that a local-moment-bearing R can serve as a knob for tuning the topological properties in quantum magnets.

RMn$_6$Sn$_6$ single crystals were synthesized via a flux method~\cite{CLATTERBUCK199978,doi:10.1080/13642819208215073}.
All these metals are magnetically ordered above room temperature (Fig.~\ref{fig:1}(e)) while the magnetic R sublattice tends to develop an anti-parallel configuration with respect to Mn if the Mn lattice is FM ordered~\cite{MALAMAN1999519, CLATTERBUCK199978}.
The magnetic anisotropy varies from easy-$ab$-plane for R = Gd, to easy-$c$-axis for R = Tb, and to a conical magnetic structure for R = Dy and Ho~\cite{VENTURINI199135,CLATTERBUCK199978,MALAMAN1999519}.
When R is Er and Tm, the Mn and Er/Tm sublattices are independently ordered in an AFM manner because the strength of the magnetic coupling is weak~\cite{MALAMAN1999519,CLATTERBUCK199978}.
As there is no $4f$ moment in LuMn$_6$Sn$_6$, its magnetic structure was reported to be a flat spiral AFM~\cite{venturini1993magnetic}, similar to that in YMn$_6$Sn$_6$~\cite{ghimire2020competing,PhysRevB.103.014416,li2020spin}.

One remarkable feature is those demonstrating FIM order (R = Gd to Er) all exhibit Shubnikov-de Haas quantum oscillations (SdH QOs) with a close and small oscillatory frequency (Fig.~\ref{fig:1}(f)).
We obtain the Berry phases according to the Lifshitz-Onsager quantization rule~\cite{shoenberg2009magnetic},
$F/B_N+\gamma=N$, where $F$ is the oscillatory frequency, $B_N$ is the N-th minimum in $\rho_{xx}$, $\gamma=1/2-\beta$, and $\beta$ is the Berry phase.
The intercepts on the N-index axis in Fig.~\ref{fig:1}(g) all give $\gamma$ close to $-1/8$, pointing to a same nontrivial Berry phase~\cite{PhysRevLett.117.077201,PhysRevB.85.033301,murakawa2013detection}.
Corresponding cyclotron mass $m^*$ is estimated to be around $0.1\,m_e$ by fitting to the Lifshitz-Kosevich formula~\cite{shoenberg2009magnetic} (Table~\ref{t1}, Section VI of the Supplemental Material~\cite{Supplemental_Material}).
The Fermi velocity ($\nu_F\simeq\sqrt{2e\hbar F}/m^*$) is found to be around $5\times10^5 \mathrm{m/s}$.

\begin{figure*}[htbp]
	\begin{center}
		\includegraphics[clip, width=0.93\textwidth]{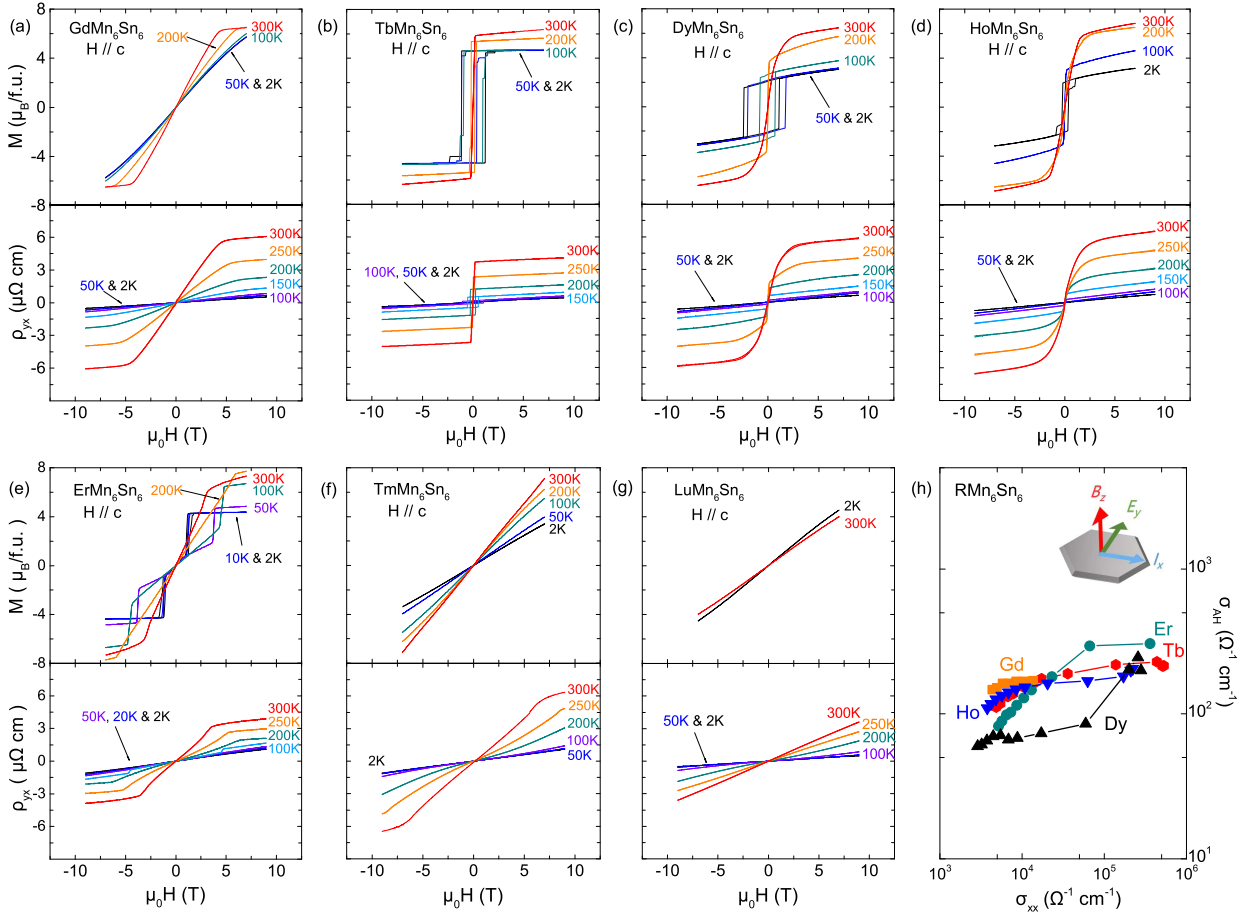}\\[1pt]
		\caption{(a)-(g): Magnetization and Hall resistivity of RM$_6$Sn$_6$. 
(h) Anomalous Hall conductivity $\sigma_{AH}$ versus $\sigma_{xx}$. Inset: Hall measurement configuration.}
		\label{fig:2}
	\end{center}
\end{figure*}

The above analyses demonstrate that the electronic structures of RMn$_6$Sn$_6$ resemble each other, which is expectable as their lattice parameters differ less than $0.4$~\%.
A series of experiments including the scanning tunneling microscopy (STM) and angle-resolved photoemission spectroscopy (ARPES)~\cite{yin2020discovery} connect the SdH QOs with the MDF in TbMn$_6$Sn$_6$.
This connection is also experimentally established in GdMn$_6$Sn$_6$ (ARPES data in Section II of the Supplemental Material~\cite{Supplemental_Material}).
We conclude that the Chern gapped MDF is generally hosted in FM Mn kagome lattice of RMn$_6$Sn$_6$ (R = Gd to Er).
We highlight two key features of these QOs that constrain our analysis along this direction.
Firstly, the Fermi surface of the pocket is detected to be 2D in angular dependent measurements~\cite{yin2020discovery}. Secondly, the QOs are intimately related with the FM order in the Mn kagome lattice (R = Gd - Ho and Er) while no QO is observed in AFM Tm, Lu or Y members.

\begin{table}
\caption{MDFs in RMn$_6$Sn$_6$.
$\sigma^{int}$: the intrinsic anomalous Hall conductivity. $F$ and $m^\ast$: frequency and cyclotron mass of SdH QOs. $v_F$ and $k_F$: Fermi velocity and wave vector.}
\begin{ruledtabular}
\begin{tabular}{p{0.9cm}<{\centering}p{1.5cm}<{\centering}p{1cm}<{\centering}p{1.1cm}<{\centering}p{1.9cm}<{\centering}p{1.4cm}<{\centering}}
R  &  $\sigma^{int}$($e^2/h$) & F(T) & m$^*$(m$_e$) & $v_F$($10^5 \mathrm{m/s}$) & $k_F$($\mathrm\AA^{-1})$ \\
\hline
Gd &  0.14 & 87 & 0.11 & 5.4 & 0.051\\
Tb &  0.12 & 96 & 0.14 & 4.5 & 0.054\\
Dy &  0.06 & 71 & 0.10 & 5.4 & 0.046\\
Ho &  0.09 & 55 & 0.12 & 4.0 & 0.041\\
Er &  0.04 & 35 & 0.11 & 3.4 & 0.033\\
\end{tabular}
\end{ruledtabular}
\label{t1}
\end{table}

To further understand the interplay between the magnetic structure and topological property, we systematically study the magnetization and anomalous Hall effect as shown in Fig.~\ref{fig:2}.
We take DyMn$_6$Sn$_6$ as an example, which has a conical magnetic structure below room temperature.
The $M(H)$ shows a hard-magnet-like profile with sharp, considerable hysteresis loops below 100~K (Fig.~\ref{fig:2}(c)).
With increasing external field, there exists a continuous rotation of the magnetic moments towards $c$-axis.
The Hall resistivity ($\rho_{yx}$) resembles the profile of its $M(H)$ completely.
We separate the anomalous contribution (Section III of the Supplemental Material~\cite{Supplemental_Material}) using the empirical relation~\cite{RevModPhys.82.1539},
\begin{equation}\label{eqn:1}
\rho_{yx}= \rho_{OH}(B)+\rho_{AH}(M)= R_0B+ R_S4\pi M
\end{equation}
where $\rho_{OH}$ and $\rho_{AH}$ represent ordinary and anomalous Hall resistivity, $R_0$ and $R_S$ are ordinary and anomalous Hall coefficients, respectively.
Similar phenomena hold for Gd, Tb and Ho members.
ErMn$_6$Sn$_6$ lies between FIM and AFM states and a small field triggers a metamagnetic transition which represents as a sharp jump on the $M(H)$ curves below 100~K (Fig.~\ref{fig:2}(e))~\cite{CLATTERBUCK199978,zhang2005unusual}.
Its $\rho_{yx}$ curves also follow the $M(H)$ curves (Fig.~S5(e)) and we can distinguish $\rho_{AH}$ as the step on the transition.
For Tm and Lu, the stable AFM structure~\cite{CLATTERBUCK199978,MALAMAN1999519} defies explicit zero-field anomalous effect in their $\rho_{yx}$, similar as the observation for YMn$_6$Sn$_6$~\cite{ghimire2020competing,PhysRevB.103.014416}.
Interestingly, $\rho_{AH}$ in TmMn$_6$Sn$_6$ deviates from the linear $M(H)$ above 100~K (Fig.~\ref{fig:2}(f)), which might arise from a topological Hall effect (THE) and deserves attention in future.


The scaling relation between the anomalous Hall conductivity (AHC) $\sigma_{AH}\simeq\rho_{AH}/\rho_{xx}^2$ and $\sigma_{xx}$ is shown in Fig.~\ref{fig:2}(h).
Previous studies have identified three distinct regimes which are delimited by $ \sigma_{xx}$ and characterized by the dependence of $\sigma_{AH}$ and $\sigma_{xx}$~\cite{RevModPhys.82.1539,PhysRevLett.97.126602,PhysRevLett.99.086602,PhysRevB.77.165103}.
For RMn$_6$Sn$_6$, $\sigma_{xx}$ ranges from $3\times 10^{3}$ to $5\times 10^{5}$ $\Omega^{-1}$ $\mathrm{cm}^{-1}$, approximately lying within the good-metal regime.
Their $\sigma_{AH}$, of the order 100~$\Omega^{-1}\mathrm{cm}^{-1}$, keeps nearly invariant in this regime, suggesting dominant intrinsic contributions.
By fitting the scaling law (Section V of the Supplemental Material~\cite{Supplemental_Material}), we extract the intrinsic Hall conductivity $\sigma^{int}$ to be around $0.14 - 0.04 e^2/h$ per kagome layer for R = Gd to Er, as summarized in Table~\ref{t1}.

\begin{figure}[htbp]
	\begin{center}
		\includegraphics[clip, width=0.49\textwidth]{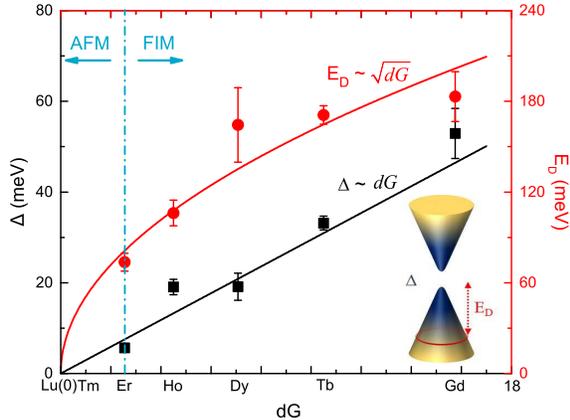}\\[1pt] 
		\caption{Derived Dirac cone energy $E_D$ and gap size $\Delta$ for the whole series. The systematic evolution of $E_D$ and $\Delta$ follows $\sqrt{dG}$ and $dG$, respectively. Inset: A sketch of a Chern gapped Dirac cone.}
		\label{fig:3}
	\end{center}
\end{figure}

A high-temperature quantum anomalous Hall effect with a quantized Hall value $e^2/h$ is theoretically supported by a Chern insulating gap of the MDF~\cite{PhysRevLett.115.186802}.
In RMn$_6$Sn$_6$, the Chern gap is near the Fermi level and therefore its Berry curvature field should induce a large intrinsic AHC.
The AHC for MDF is estimated as~\cite{PhysRevB.75.045315}
\begin{equation}\label{eqn:2}
\sigma^{int}=\frac{e^2}{h}\frac{\Delta}{\sqrt{\Delta^2+4\hbar^2 k_F^2 v_F^2}}=\frac{e^2}{h}\frac{\Delta/2}{E_D}.
\end{equation}
As $\Delta$ and $E_D$ in TbMn$_6$Sn$_6$ have been completed depicted by STM, the agreement between theoretically inferred AHC and our transport data indicates that the MDF dominates the Berry curvature field~\cite{yin2020discovery}.
Assuming $\sigma^{int}$ is stemming from the MDF in the kagome lattice in other RMn$_6$Sn$_6$ as well, we evaluate the effective $\Delta$ and $E_D$.
Interestingly, both parameters gradually decrease when R changes from Gd to Er (Fig.~\ref{fig:3}).

The analysis highlights the possibility of Chern phase engineering in RMn$_6$Sn$_6$ by changing the R atoms.
Regulating the Chern gap has been less accomplished in experiments and one example is Fe$_3$Sn$_2$ whose gap can be modulated by applying an external magnetic field~\cite{yin2018giant,ye2019haas}.
Here we observe a unique, intrinsic R engineering of topological electrons in a kagome magnet, and its mechanism needs further elaboration.
The chemical pressure does not likely play a role because the lattice parameter changes little in the series.
On the other hand, the total SOC strength should increase slightly with the atomic number from R = Gd to Er, which is opposite to the trend of gap closing.
Below we discuss one possible mechanism.

\begin{figure}[htbp]
	\begin{center}
		\includegraphics[clip, width=0.4\textwidth]{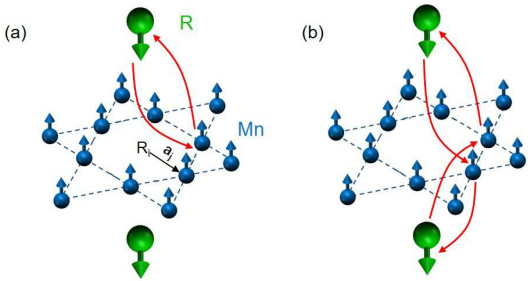}\\[1pt] 
		\caption{A sketch of possible electron hopping between R and Mn kagome in FIM RMn$_6$Sn$_6$. (a) First-order process involving one-electron hopping.
			(b) Second-order process involving two-electron hopping.
		}
		\label{fig:4}
	\end{center}
\end{figure}

When we plot $\Delta$ and $E_D$ against $dG = (g_J-1)^2J (J+1)$, where $g_J$ is the Land\'e factor and $J$ is the total angular momentum of the $R^{3+}$ ion Hund's rule ground state,
we find they apparently follow a linear and square root relation, respectively, as shown in Fig.~\ref{fig:3}.
We propose that the coupling between $4f$ moments and $3d$ electrons of Mn plays a crucial role in this engineering.
If we include the hopping between Mn and R (Fig.~\ref{fig:4}(a)), it will introduce an additional diagonal component in total Hamiltonian, which is proportional to $-J_Hm$, where $J_H$ is the Hund's coupling, $m$ equals $(g_J-1)\sqrt{J (J+1)}=\sqrt{dG}$ is the effective moment of R ions in mean-field approximation~\cite{jensen1991rare}.
Considering this first-order hopping process we naturally conclude that the chemical potential of the MDF should shift linearly with $\sqrt{dG}$.
The first-order process should not contribute to the gap opening on $K(K^\prime)$ valley because of the symmetry of a kagome lattice.
To elucidate the gap opening, higher order process (for instance, the two-electron hopping in Fig.~\ref{fig:4}(b)) has to be considered (Section VII of the Supplemental Material~\cite{Supplemental_Material}).
The forth order perturbation will open a gap proportional to $\sim J_H^2m^2$, namely $\Delta\propto dG$, just as what we have observed in experiment.

The magnetic exchange coupling between $3d$ and $4f$ electrons plays an important role in the magnetic properties of RMn$_6$Sn$_6$ as well~\cite{zhang2005unusual,VENTURINI199135,CLATTERBUCK199978,MALAMAN1999519}.
Phenomenologically, this coupling is weakened when R changes from Gd to Tm.
ErMn$_6$Sn$_6$ has a critically weak coupling which makes it lie at the border between FIM and AFM ground states and we notice it also has the smallest $\Delta$ and $E_D$.
This interesting observation supports the critical role of magnetic coupling in the MDF structure.
On the other hand, the THE and electronic structure in YMn$_6$Sn$_6$ which has no 4f moments, deserve further elaborations~\cite{ghimire2020competing,PhysRevB.103.014416,li2020spin}.
Our work highlights the material system which not only largely extends the family of quantum kagome magnets, but also intrinsically contains a quantum knob for Chern phase engineering.
The quantum tuning of the topological band structures via magnetic coupling may serve as a pathway to engineer the topological phases in correlated systems.
\\
\section*{Acknowledgments}
We thank Prof. Weiwei Xie for helpful discussions about the crystal structure and Jingzhuo Zhou for assistance in figure drawing.
This work was supported by the National Natural Science Foundation of China No.U1832214, No.11774007, No.U2032213, No.11774352, the National Key R \& D Program of China (2018YFA0305601) and the strategic Priority Research Program of Chinese Academy of Sciences, Grant No. XDB28000000.
X. Xu acknowledges support from the China Postdoctoral Science Foundation No.2020M682056 and Anhui Postdoctoral Foundation No.2020B472.
X. Xu is also supported by CAS Key Laboratory of Photovoltaic and Energy Conservation Materials Foundation No.PECL2020ZZ007 and Special Research Assistant, Chinese Academy of Sciences.
Experimental and theoretical work at Princeton University was supported by the Gordon and Betty Moore Foundation (Grant No. GBMF4547 and GBMF9461/Hasan).
Work at Princeton University and Princeton-led synchrotron-based ARPES measurements are supported by the U.S. Department of Energy under Grant No. DOE/BES DE-FG-02-05ER46200. Z.C. thank D. Lu and M. Hashimoto for beamtime support at SSRL Beamline 5-2.

%

%
%

\clearpage

\end{document}